\documentclass[doublecol]{epl2} 
\usepackage{xcolor,amsmath,amssymb}
\usepackage{marginnote}

\shorttitle{Taxis of cargo-carrying microswimmers in traveling activity waves} 

\title{Taxis of cargo-carrying microswimmers in traveling activity waves} 

\author{Pietro Luigi Muzzeddu \inst{1,2}   \and \'Edgar Rold\'an\inst{2} \and Andrea Gambassi\inst{1,3} \and Abhinav Sharma\inst{4,5}}
\shortauthor{P. L. Muzzeddu \etal}

\institute{                    
  \inst{1} SISSA - International School for Advanced Studies, via Bonomea 265, 34136 Trieste, Italy\\
  \inst{2} ICTP - The Abdus Salam International Centre for Theoretical Physics, Strada Costiera 11, 34151 Trieste, Italy\\
  \inst{3} INFN, Sezione di Trieste, Trieste, Italy\\
  \inst{4} Leibniz-Institut f\"ur Polymerforschung Dresden, Institut Theorie der Polymere, 01069 Dresden, Germany\\
  \inst{5} Technische Universit\"at Dresden, Institut f\"ur Theoretische Physik, 01069 Dresden, Germany
}
\pacs{05.70.Ln}{Nonequilibrium and irreversible thermodynamics}

\abstract{%
Many fascinating properties of biological active matter crucially depend on the capacity of constituting entities to perform directed motion, e.g., molecular motors transporting vesicles inside cells or bacteria searching for food. While much effort has been devoted to mimicking biological functions in synthetic systems, such as transporting a cargo to a targeted zone, theoretical studies have primarily focused on single active particles subject to various spatial and temporal stimuli. Here we study the behavior of a self-propelled particle carrying a passive cargo in a
travelling activity wave and show that this active-passive dimer displays a rich, emergent tactic behavior.  
For cargoes with low mobility, the dimer always drifts in the direction of the wave propagation. For highly-mobile cargoes, instead, the dimer can also drift against the traveling wave. The transition between these two tactic behaviors is controlled by the ratio between the frictions of the cargo and the microswimmer. In slow activity waves the dimer can perform an \emph{active surfing} of the wave maxima, with an average drift velocity equal to the wave speed. These analytical predictions, which we confirm by numerical simulations, might be useful for the future efficient design of bio-hybrid microswimmers. %
}

\begin{document}

\maketitle

\section{Introduction}

The ability to self-propel at the expense of  fuel consumption
is a fundamental property of active matter~\cite{julicher1997modeling,hanggi2009artificial,marchetti2013hydrodynamics,bechinger2016active}. 
In the biological context, self-propelling
microscopic systems perform functions that require accurate
directed transport, for instance, white blood cells chase intruders~\cite{fenteany2004cytoskeletal}, motor proteins transport RNA inside cells~\cite{kanai2004kinesin} and microswimmers such as 
E.~coli~\cite{berg2004coli} and sperm cells~\cite{friedrich2007chemotaxis} steer themselves towards sources of nutrients. Directed transport is a highly desirable
property, in particular for applications in drug delivery at the nanoscale~\cite{mano2017optimal,reinivsova2019micro,ebbens2016active, Garcia2013micromotor,Sanchez2014chemically}. For this purpose, 
bio-hybrid microswimmers have been designed by integrating biological entities with synthetic constructs, e.g., bacteria capable to transport and drop off  passive microscopic cargo to specific target locations~\cite{singh2017microemulsion,alapan2018soft,vaccari2018cargo,senturk2020red}. 

Bacteria and eukaryotic cells~\cite{Fisher1989-lb,MARTIEL1987807} generally navigate  in dynamic activating media and react {\em in vivo} to 
time-dependent tactic stimuli of various nature. Such an interaction with travelling activity signals, e.g., chemical waves~\cite{Tomchik1981}, leads to fascinating collective behavior~\cite{Gregor2010_onset} and sometimes to unexpected migration phenomena, as in the case of the so-called \emph{chemotactic wave paradox}~\cite{Tomchik1981,HOFER19941}. 
While synthetic active particles mimic the basic features of self-propulsion and persistence of actual biological active matter, they lack the information processing capacity and motoric control which is essential for directed transport in biological and bio-hybrid systems. 
Despite their memory-less response to tactic signals, artificial self-propelled particles exhibit directed transport when immersed in travelling  waves controlling locally their degree of activity (e.g. their self propulsion velocity), as  shown experimentally with phototactic Janus particles exposed to propagating optical pulses~\cite{lozano2019diffusing}. 
Several theoretical studies have focused on controlling and directing the motion of a single self-propelled particle in a fluctuating environment~\cite{geiseler2016chemotaxis, geiseler2017selfpolarizing, geiseler2017artificial, sharma2017brownian,merlitz2018linear}. However, a fundamental understanding of the behavior of cargo-carrying microswimmers in time-dependent activity is still lacking.

Cargo-carrying self-propelled particles have been analyzed in a stationary, but spatially inhomogeneous activity~\cite{vuijk2021chemotaxis}. While a single self-propelled particle always accumulates in regions with low activity, attaching a passive cargo reverses this tendency. In fact, beyond a certain threshold cargo, the particle accumulates in regions with larger activity~\cite{vuijk2021chemotaxis}. 
While preferential accumulation could be regarded as a signature of the tactic behavior \cite{vuijk2021chemotaxis}, in the case of stationary activity, it causes no transport of the dimer. 
By contrast, for a time-dependent activity, such as a source emitting activity pulses, the tactic behavior of an active particle can result in motion towards or away from the source.
With this motivation, in this paper we study active-passive dimers subject %
to a time-dependent activity in the form of a travelling wave. We analytically show that the dimer exhibits directed transport, characterized by a wave-induced drift. The direction of this drift depends on the wave speed, being opposite to its propagation direction for a slow wave but along it for a fast wave. Interestingly, the opposite drift vanishes at a threshold cargo upon increasing its friction, beyond which the dimer only shows drift along the propagation direction. We show that the threshold value of the cargo coincides with that existing in the stationary activity. 
Our theoretical treatment of the active-passive dimer is based on the active Ornstein-Uhlenbeck particle (AOUP) model of activity~\cite{caprini2019comparative,martin2021statistical,caprini2022parental,gopal2021energetics}. Our analysis shows that the AOUPs are completely equivalent to active Brownian particles \cite{vuijk2021chemotaxis}  (ABPs) in terms of their tactic behavior.

\section{The model}
\label{The_model}

\begin{figure}[t]
\begin{center}
\includegraphics[width=0.49\textwidth]{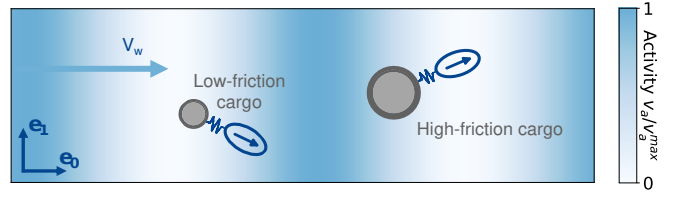}
\caption{Sketch of the stochastic model described by eqs.~\eqref{eq:model} in two spatial dimensions.
A self-propelled active microswimmer (blue ellipse) in a fluid drags a passive cargo (gray circle)   via a harmonic interaction (blue spring). The instantaneous self-propulsion velocity of the microswimmer (blue arrow) is  locally controlled  by a sinusoidal traveling wave of activity, which   propagates through the fluid with phase velocity $v_w$ along the unit vector $\bm{e}_0$.  
For illustration we sketch here two examples of active-passive dimers, one with a low-friction cargo ($q$ small, left) and the other with a high-friction cargo ($q$ large, right).
}
\label{sketch}
\end{center}
\end{figure}

In this section we introduce a minimal model for the dynamics of an active microswimmer dragging a passive load in $d$ spatial dimensions within an 
inhomogeneous and time-dependent environment. The microswimmer at position $\bm{r}$ and time $t$ interacts with a tactic signal described by the activity field $v_{\rm a}(\bm{r}-\bm{v}_wt)$, which propagates with velocity $\bm{v}_w = v_w\bm{e}_0$ along the direction of the unit vector $\bm{e}_0$, as depicted in fig.~\ref{sketch}.
As usually done for $\mu$m-sized colloidal particles in a liquid, we assume that viscous forces dominate over inertial effects and therefore we consider an overdamped dynamics for the active-passive dimer, which is governed by the following Langevin equations:
\begin{subequations}
\label{eq:model}
\begin{eqnarray}
       && \hspace{-1cm}\dot{\bm{r}}_1=-\frac{1}{\gamma}\nabla_{\bm{r}_1} U(\bm{r}_1-\bm{r}_2)+v_{\rm a}(\bm{r}_1-\bm{v}_wt)\bm{\eta} + \sqrt{2D}\bm{\xi}_1,\label{dynamics1}\\
        &&\hspace{-1cm}\dot{\bm{r}}_2=-\frac{1}{q\gamma}\nabla_{\bm{r}_2} U(\bm{r}_1-\bm{r}_2) + \sqrt{\frac{2D}{q}}\bm{\xi}_2, \label{dynamics2} \\
        &&\hspace{-1cm}\tau \dot{\bm{\eta}}=-\bm{\eta}+\sqrt{\frac{2\tau}{d}}\bm{\xi}_3;\label{dynamics3}    
\end{eqnarray}
\end{subequations}
where $\bm{r}_1$ and $\bm{r}_2$ denote the positions of the active microswimmer and the passive cargo, respectively. The interaction $U(\bm{r}_1-\bm{r}_2)$ between them is modeled by an isotropic parabolic potential $U(\bm{r})=\kappa \bm{r}^2/2$, with stiffness $\kappa>0$ and zero rest length. 
The stochastic forces $\bm{\xi}_1$, $\bm{\xi}_2$, $\bm{\xi}_3$ are three independent zero-mean Gaussian white noises accounting for thermal fluctuations. Moreover, the active carrier exploits local energy injections to self-propel along the direction of the propulsion vector $\bm{\eta}$ 
which is given by a set of $d$ independent Ornstein-Uhlenbeck processes with variance $1/d$ and correlation time $\tau$. It follows that $\bm{\eta}$ is a zero-mean Gaussian colored noise with autocorrelation function~$\left\langle \eta_\alpha(t) \eta_\beta(s) \right\rangle=(\delta_{\alpha,\beta}/d)\exp\left(-|t-s|/\tau\right)$, where $\delta_{\alpha,\beta}$ denotes Kronecker's delta.
This normalization ensures that the average modulus squared of the propulsion vector is $\left< \lVert \bm{\eta} \rVert^2\right>=1$ for all values of $d$.  
While the time scale $\tau$ sets the persistence of the self-propulsion force, its strength is modulated in space by the activity field $v_{\rm a}$. In order to recover an equilibrium dynamics in the absence of activity $v_{\rm a}=0$, we connect the mobility $\gamma$ and the diffusivity $D$ via the Einstein relation $D=k_{\rm B}T/\gamma$. Moreover, the cargo and the active carrier are assumed to have different friction coefficients, the ratio of which is given by the parameter $q$. In a Newtonian fluid and for spherical colloidal carrier and cargo, $q$ equals the ratio of the radius of the cargo to that of the carrier.

The Langevin dynamics in eqs.~\eqref{eq:model} can be more conveniently written in terms of the dimer position in the comoving frame, which we identify with the centre of friction $\bm{\chi}=(\bm{r}_1+q\bm{r}_2)/(1+q)-\bm{v}_wt$
and the distance $\bm{r}=\bm{r}_1-\bm{r}_2$. Changing variables $(\bm{r}_1,\bm{r}_2, \bm{\eta})\to (\bm{\chi},\bm{r},\bm{\eta})$ to the new coordinate system, the Fokker-Planck equation for the probability density $P(\bm{\chi},\bm{r},\bm{\eta},t)$ associated to the stochastic dynamics in eqs.~\eqref{eq:model} reads: 
\begin{equation}
\begin{split}
    &\partial_t P(\bm{\chi},\bm{r},\bm{\eta},t)=\frac{1}{d\tau}\hat{\mathcal{L}}_{\bm{\eta}} P\,+\\&-\nabla_{\bm{\chi}} \cdot \left[ -\bm{v}_w P +  \frac{1}{1+q}v_{\rm a}\left(\bm{\chi}'\right)\bm{\eta}P - \frac{D}{1+q}\nabla_{\bm{\chi}}P \right]+\\
    &-\nabla_{\bm{r}}\cdot \left[ -\frac{1+q}{q\gamma} \nabla_{\bm{r}} UP+v_{\rm a}\left(\bm{\chi}'\right)\bm{\eta}P - \frac{1+q}{q}D \nabla_{\bm{r}} P \right]\,,
\end{split}
\end{equation}
with $\bm{\chi}'=\bm{\chi}+q\bm{r}/(1+q)$ and $\hat{\mathcal{L}}_{\bm{\eta}} P=\nabla^2_{\bm{\eta}}P+d \nabla_{\bm{\eta}} \cdot \left(\bm{\eta} P \right)$.

\section{Transport properties for slow activity waves} 
In order to estimate the extent to which the propagating tactic signal affects the directed motion of the cargo-carrying microswimmer, we focus on transport properties induced by the activity travelling wave. 
With the help of a mean-field hydrodynamic theory, we derive an effective dynamics which describes the evolution of the dimer at time scales longer than $\tau$ and length scales larger than the persistence length $l_p\sim v_{\rm a}\tau$ \cite{cates2013when}. In particular, the predictions deriving from such hydrodynamic theory are expected to be valid for activity fields which are slowly varying on the length scale $l_p$ ({\em large wavelength} approximation). In order to identify all relevant hydrodynamic variables, i.e., those fields the relaxation time of which grows indefinitely upon increasing the wavelength (slow modes), we perform a moment expansion analogous to, e.g., refs.~\cite{cates2013when, Solon2015comp,Adeleke2020}.

The evolution of the modes is described by a hierarchical structure, the detailed derivation of which is reported in sec.~1 of Supplementary Material (SM). Importantly, we note that the zeroth order mode $\varphi(\bm{\chi},\bm{r},t)= \int d\bm{\eta}\,P(\bm{\chi},\bm{r},\bm{\eta},t)$, which describes the density related to the spatial marginal variables $\bm{\chi}$ and $\bm{r}$, is the only slow mode of the system. Indeed, $\varphi(\bm{\chi},\bm{r},t)$ is associated with a conservation law and its dynamics has the form of a continuity equation:
\begin{equation}
\begin{split}
    &\partial_t \varphi(\bm{\chi},\bm{r},t)= -\partial_\alpha \left[-v_w \delta_{\alpha,0}\varphi + \frac{v_{\rm a}\left(\bm{\chi}'\right) \sigma_\alpha}{(1+q)} -\frac{D}{1+q}\partial_\alpha \varphi \right]\\
        &\quad-\partial'_\alpha \left[ -\frac{(1+q)}{q\gamma} \partial'_\alpha U \varphi +v_{\rm a}\left(\bm{\chi}'\right)\sigma_\alpha - \frac{(1+q)D}{q} \partial'_\alpha \varphi \right],
        \end{split}
    \label{mode0}
\end{equation}
where we introduced the shorthand notation 
$\partial_\alpha \equiv \partial_{\chi_\alpha}$ and 
$\partial'_\alpha \equiv  \partial_{r_\alpha}$,
%
while repeated indices imply summation. Furthermore, $\sigma_\alpha$ is the $\alpha$-th component of the first-order mode $\bm{\sigma}(\bm{\chi},\bm{r},t)=\int d\bm{\eta}\,\bm{\eta}P(\bm{\chi},\bm{r},\bm{\eta},t)$, which is related to the conditional average polarization at fixed spatial variables. Its dynamics is governed by
\begin{equation}
    \begin{split}
      &\partial_t\sigma_\alpha(\bm{\chi},\bm{r},t)= -\frac{\partial_\alpha \left[ v_{\rm a}\left(\bm{\chi}'\right)\varphi \right]}{(1+q)d}-\frac{\partial'_\alpha \left[v_{\rm a}\left(\bm{\chi}'\right)\varphi\right]}{d}\\
      &\quad\quad\quad+\frac{(1+q)}{q\gamma}\partial'_\beta \left[  \partial'_\beta U \sigma_\alpha \right] - \tau^{-1}\sigma_\alpha + \mathcal{O}(\partial^2)\,,
    \end{split}
    \label{mode1}
\end{equation}
where dependencies on higher-order modes are included in $\mathcal{O}(\partial^2)$. Notably, the decay rate due to the sink term $- \tau^{-1}\sigma_\alpha$ makes $\sigma_\alpha(\bm{\chi},\bm{r},t)$ a fast mode which does not obey a conservation law and which can be described by a quasi-static approximation. Moreover, the contribution $\mathcal{O}(\partial^2)$ of higher-order gradients is negligible under the assumption of a slowly varying activity field. 

%
%
\begin{figure}[t]
\begin{center}
\includegraphics[scale=0.3]{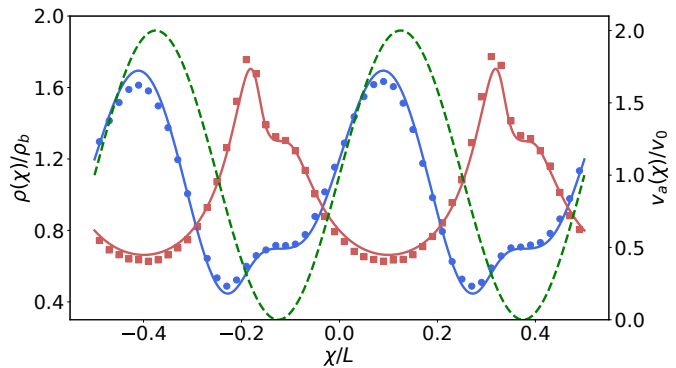}
\caption{Stationary density $\rho(\chi)$ of the dimer (left axis), 
in the comoving frame of the traveling activity wave $v_{\rm a}(\chi)$ with sinusoidal shape (green dashed line, eq.~\eqref{activity_field}, right axis), as obtained from 
numerical simulations (symbols) and from analytical predictions (eq.~\eqref{steady_state_density}, solid lines).  
The latter hold under the assumption of long wavelength and slow traveling wave and they are reported for both a high-friction cargo with $q = q_{\rm high} >q_{\rm th}$ (blue) and a low-friction  cargo with $q = q_{\rm low} <q_{\rm th}$ (red). 
Both analytical and numerical predictions have been obtained by assuming periodic boundary conditions. The numerical data  were obtained  from a single Langevin-dynamics simulation  of duration $10^8$ using Euler-Maruyama scheme with time step $\Delta t=0.01$. Other simulation parameters: $v_w=10^{-2}$, $v_0=1.0$, $\tau=0.1$, $\kappa=5$, $\gamma=1.0$, $D=10^{-3}$, $\lambda=10/(4\pi)$, $q_{\rm high}=4$ and $q_{\rm low}=1$.
}
\label{stationary_density}
\end{center}
\end{figure}

The combination of large-wavelength approximation and quasi-stationarity of $\bm{\sigma}(\bm{\chi},\bm{r},t)$ at time scales longer than $\tau$ provides a closure scheme for the hierarchy without needing information about higher-order modes. In particular, after integrating out the relative coordinate $\bm{r}$, we derive an effective drift-diffusion equation for the marginal density  $\rho(\bm{\chi},t)=\int d\bm{r}\varphi(\bm{\chi},\bm{r},t)$ (see sec.~2 of SM for the detailed derivation), which reads:
\begin{equation}
    \partial_t \rho(\bm{\chi},t)=-\nabla_{\bm{\chi}}\cdot \left[\bm{V}_{\rm eff}(\bm{\chi})\rho(\bm{\chi},t) - \nabla_{\bm{\chi}}(D_{\rm eff}(\bm{\chi})\rho(\bm{\chi},t)) \right],
    \label{drift-diff}
\end{equation}
where the effective drift and effective diffusivity are given, respectively, by
\begin{eqnarray}
   \bm{V}_{\rm eff}(\bm{\chi})&=&(1-\epsilon/2)\nabla_{\bm{\chi}} D_{\rm eff}(\bm{\chi}) -\bm{v}_w,\\
   D_{\rm eff}(\bm{\chi})&=&\frac{D}{1+q}+\frac{\tau v^2_{\rm a}(\bm{\chi})}{d(1+q)^2} \,.
   \label{effective_terms}
\end{eqnarray}
This expression of $D_{\rm eff}$ reveals an enhancement of the diffusivity $D/(1+q)$ of the center of friction induced by the activity via a term $\propto v^2_{\rm a}(\bm{\chi})$. 
Interestingly, the  alignment of the effective drift with the activity gradient is controlled  by the {\em tactic coupling} 
\begin{equation}
    \epsilon=1-\frac{q}{1+\frac{1+q}{q}\frac{\tau}{\tau_{\rm r}}}\,,
    \label{epsilon_coupling}
\end{equation}
where $\tau_{\rm r}=\gamma/\kappa$ is the characteristic spring relaxation time. 
The role of $\epsilon$ can be understood by considering the case of static activity field. In fact, for $v_w=0$, the stationary density obtained from eq.~\eqref{drift-diff} is 
\begin{equation}
    \rho(\bm{\chi}) = \mathcal{N}^{-1}\left[ 1+\frac{\tau v^2_{\rm a}\left(\bm{\chi}\right)}{dD(1+q)} \right]^{-\epsilon/2}.
    \label{eq:densityst}
\end{equation}
Accordingly, $\epsilon$ determines the preferential accumulation of the dimer in the regions with high or low activity depending on 
its sign. Here, $\mathcal{N}$ is a normalization constant. 
Equation~\eqref{epsilon_coupling} implies that for a fixed $\tau/\tau_{\rm r}$, the tactic coupling $\epsilon$  is entirely determined by the friction ratio $q$, because it changes sign at the threshold value 
\begin{equation}
    q_{\rm th}=\frac{1}{2}\left[1+\tau/\tau_{\rm r}+\sqrt{\left(1+\tau/\tau_{\rm r} \right)^2 + 4\tau/\tau_{\rm r}}\right] \ge 1.
    \label{q_threshold}
\end{equation}
For highly mobile cargoes with $q<q_{\rm th}$ one has $\epsilon>0$ and thus the dimer preferentially accumulates in low-activity regions. For slow cargoes with $q>q_{\rm th}$, instead, $\epsilon<0$ and the dimer preferentially accumulates in high-activity regions.
Interestingly, as in the single-particle case (see, e.g., ref.~\cite{caprini2022dynamics}), the equivalence with a cargo-carrying ABP~\cite{vuijk2021chemotaxis} with rotational diffusivity $D_r$ is fully recovered by imposing $\tau^{-1}=(d-1)D_r$.

%
%
\begin{figure}[t]
\begin{center}
\includegraphics[scale=0.4]{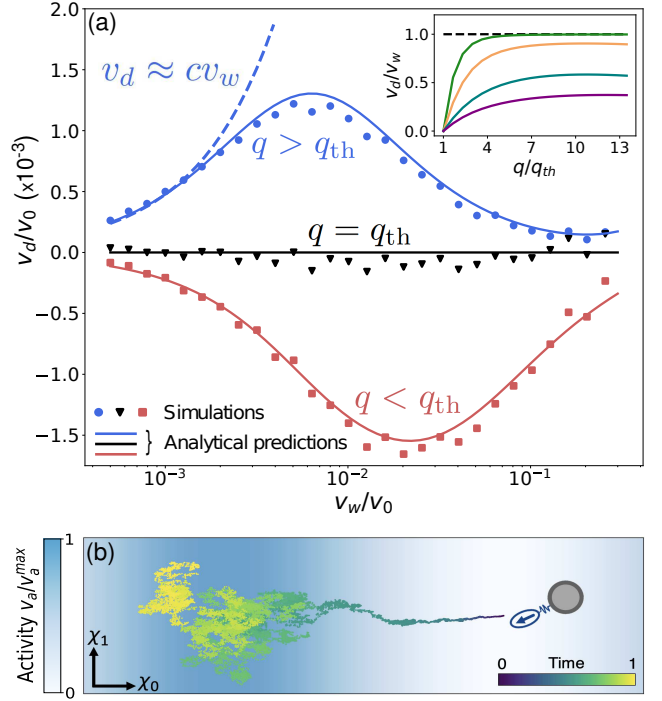}
\caption{
(a) Average drift $v_d$ as a function of the phase velocity $v_w$ in the slow-wave regime $v_w<v_0$ (eq.~\eqref{vd_expression}). For low-friction cargoes with $q=q_{\rm low}<q_{\rm th}$ (red line), the microswimmer exhibits a negative tactic behavior. At the threshold value $q_{\rm th}$ (black line), the average drift vanishes for all wave velocities $v_w$, whereas for $q=q_{\rm high}>q_{\rm th}$ (blue solid line),
the dimer is characterized by positive taxis.  
Numerical results (symbols) have been obtained by computing the quantity $(\chi(t)+v_w t -\chi(0))/t$ for each of the $N=10^3$ independent stochastic trajectories of length $t$, and averaging over different realizations. The remaining simulation parameters are $v_0=1.0$, $\tau=0.1$, $\kappa=5$, $\gamma=1.0$, $D=10^{-2}$, $\lambda=10/(4\pi)$, $q_{\rm high}=4$ and $q_{\rm low}=1$. In the inset, we report the slope of the linear relation $v_d\approx c v_w$ (blue dashed line) at small wave velocities as a function of $q$, and for thermal diffusivity $D\in \{0.05,\,0.03,\,0.01,\,0.001 \}$ (solid lines from bottom to top).
(b) Stochastic trajectory of a cargo-carrying microswimmer in the comoving frame $(\chi_0,\chi_1)$ in two spatial dimensions. 
For a high-friction cargo $(q=20)$ and small thermal diffusivity $D=10^{-3}$ the dimer ``surfs'' the propagating activity wave by localizing around its maximum while traveling with the same velocity, i.e., $v_d=v_w$. 
}
\label{drift_small}
\end{center}
\end{figure}
%
%

In order to analyze the general case of an activity
travelling wave ($v_w\neq0$), we assume for simplicity that the activity field $v_{\rm a}$ varies only along $\bm{e}_0$.
Accordingly, we denote the effective drift and diffusivity with $D_{\rm eff}(\chi_0)$ and $V_{\rm eff,\alpha}(\chi_0)$ as they now depend only on $\chi_0=\bm{\chi}\cdot \bm{e}_0$. 
As an example, we hereafter consider the sinusoidal wave
\begin{equation}
    v_{\rm a}(\chi_0)=v_0 \left[1+\sin(\chi_0/\lambda) \right],
 \label{activity_field}
\end{equation}
with wavelength $\lambda$.
The resulting stationary density $\rho(\bm{\chi})$ in the comoving frame can be determined by considering an ensemble of non-interacting dimers with initial bulk density $\rho_b=L^{-d}$, $L$ being their typical interparticle distance. 
In this way, from eq.~\eqref{drift-diff} we find 
\begin{equation}
    \frac{\rho(\bm{\chi})}{\rho_b}=\frac{L\, D_{\rm eff}^{-1}(\chi_0) \int_0^L dx\exp \left\{-\int_{\chi_0}^{\chi_0+x}dy\frac{V_{\rm eff,0}(y)}{D_{\rm eff}(y)}  \right\}}{\int_0^L\,du\int_0^Ldx\, D_{\rm eff}^{-1}(u)\exp \left\{-\int_{u}^{u+x}dy\frac{V_{\rm eff,0}(y)}{D_{\rm eff}(y)}  \right\}},
    \label{steady_state_density}
\end{equation}
which is illustrated in fig.~\ref{stationary_density} and which also features the transition in the preferential accumulation illustrated above for $v_w=0$. 
Moreover, the interaction with a propagating activity field induces a non-trivial tactic response in the microswimmer, which is now able to sustain a 
non-vanishing stationary flux $J_0$ in the comoving frame, acquiring an average drift velocity $v_d=(\langle\dot{\bm{r}}_1\rangle +q\langle\dot{\bm{r}}_2\rangle)/(1+q) = J_0/\rho_b+v_w$ along $\bm{e}_0$ in the lab frame. 
This drift is given by~\cite{hanggi1990reaction, goel2016stochastic}
\begin{equation}
    v_d=\frac{L \left[ 1-\exp \left\{-\int_{0}^{L}dy\frac{V_{\rm eff,0}(y)}{D_{\rm eff}(y)}  \right\}\right]}{\int_0^L\,du\int_0^Ldx\, D_{\rm eff}^{-1}(u)\exp \left\{-\int_{u}^{u+x}dy\frac{V_{\rm eff,0}(y)}{D_{\rm eff}(y)}  \right\}}+v_w,
    \label{vd_expression}
\end{equation}
and it strongly depends on the tactic coupling $\epsilon$ and therefore on $q$. More precisely, it can be shown analytically that $v_d$ vanishes at the static threshold  value $q=q_{\rm th}$ in eq.~\eqref{q_threshold} (see sec.~3 of SM). Additionally, for sufficiently small thermal diffusivity $D$, the threshold value $q=q_{\rm th}$ also separates two distinct tactic regimes with respect to the wave propagation: {\em positive taxis} for $q>q_{\rm th}$,  where the microswimmer navigates along the propagating tactic signal with $v_d/v_w>0$, and {\em negative  taxis} for $q<q_{\rm th}$, where the microswimmer navigates against it, with $v_d/v_w<0$, see fig.~\ref{drift_small}(a).
This predicted negative taxis as well as the fact that its magnitude decreases upon increasing $D$
are consistent with what occurs for a single active particle \cite{geiseler2017selfpolarizing, geiseler2016chemotaxis}, which is retrieved as the limit $q\to0$ of our model. 
Conversely, when $q>q_{\rm th}$, the cargo-carrying microswimmer travels
along the sinusoidal wave due to its tendency to  localize close to the propagating activity crests, performing the \emph{active surfing} shown in fig.~\ref{drift_small}(b). 
Interestingly, an analogous effect was observed experimentally with single self-polarizing phototactic particles in traveling light pulses~\cite{lozano2019diffusing}. While in ref.~\cite{lozano2019diffusing} 
this behavior is caused by an aligning torque, in our model it emerges 
as a cooperative effect between the active carrier and the passive cargo. 
Note, however, that the ability of the
microswimmer to catch up with the travelling wave crests, i.e., $v_d\simeq v_w$
is limited to the case of slowly propagating activity wave, which explains the non-monotonicity of the blue curve in fig.~\ref{drift_small}(a). 
In order to quantify the efficiency of this surfing, we determine the slope $c$ of the linear relation $v_d\approx c v_w$, which holds at small wave velocities $v_w$.
Its dependence on $q$ and the thermal diffusivity $D$ is reported in the inset of fig.~\ref{drift_small}, which shows, as expected, that $c\leq1$ and that the directed transport is highly efficient (i.e., $c\simeq 1$) for $D\ll\tau v_0^2$.

We recall here that the predictions presented above follow
from a coarse graining which assumes
that the activity field varies slowly on a length scale of the order of  $l_p=v_0\tau$. 
In the static case $v_w=0$, this condition is met for $\lambda \gg l_p$. 
However, for a traveling wave, the coarse graining additionally 
requires that the distance $\sim v_w\tau$ 
traveled by the active wave on a time scale $\sim \tau$ does not exceed $\sim l_p$, which happens for $v_w<v_0$. 
Accordingly, in order to investigate the transport properties in the opposite case $v_w>v_0$, we pursue below an alternative analytical approach. 

\section{Transport properties for fast activity waves}
For simplicity, and without loss of generality, we restrict the analysis of the case $v_w>v_0$ to one-dimensional systems and to a sinusoidal traveling wave as in eq.~\eqref{activity_field}.
The main difference compared to the slow-wave approximation discussed above
lies in the closure scheme used to combine the mode eqs.~\eqref{mode0} and \eqref{mode1}. More precisely, as the small gradients approximation is no longer applicable for $v_w>v_0$, we explore this regime by considering small self-propulsion forces by keeping in the effective dynamics only contributions of the lowest order in $v_0$~\cite{sharma2016communication,merlitz2018linear,dal2019linear}.
To this aim, we rewrite eq.~\eqref{mode1} in the more convenient form
\begin{equation}
    \hat{\mathcal{L}}_{\sigma}\sigma(\chi,r,t)=-\frac{\partial_\chi \left[ v_{\rm a}(\chi')\varphi \right]}{(1+q)}-\partial_r \left[ v_{\rm a}(\chi')\varphi \right]+\Upsilon(\chi,r,t),
    \label{eq_pol_2}
\end{equation}
where $\chi'=\chi+qr/(1+q)$ is the position of the active carrier in the comoving frame, $\Upsilon(\chi,r,t)$ includes all contributions of 
higher-order modes, and the operator $\hat{\mathcal{L}}_{\sigma}$ is defined as
\begin{equation}
\hat{\mathcal{L}}_{\sigma}=\partial_t + \frac{1}{\tau} -v_w\partial_\chi -\frac{D}{1+q} \partial^2_\chi - \frac{(1+q)D}{q}\left[\partial^2_r + \frac{1}{\ell^2}\partial_r r\right],
\label{op_sigma}
\end{equation}
with the characteristic length $\ell=\sqrt{D \tau_{\rm r}}$.

\begin{figure}[t]
\begin{center}
\includegraphics[scale=0.28]{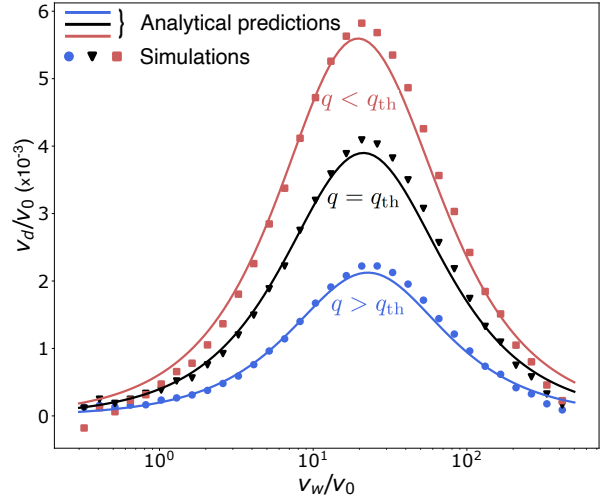}
\caption{Average drift velocity $v_d$ as a function of the phase velocity $v_w$ of the activity wave for $v_w>v_0$ (eq.~\eqref{vd_predicted}). The cargo-carrying microswimmer acquires a positive drift independently of 
the value of the friction ratio $q$, which takes here the same values as those of the corresponding curves in fig.~\ref{drift_small}.
The results from numerical simulations and analytical predictions 
have been obtained as described in the caption of fig.~\ref{drift_small}, with the same set of parameters.}
\label{drift_big}
\end{center}
\end{figure}
%
%
%

%
To solve for $\sigma(\chi,r,t)$, we then determine 
the Green function of $\hat{\mathcal{L}}_{\sigma}$ and  compute the convolution with the r.h.s.~of eq.~\eqref{eq_pol_2}. In doing this, we assume that the contribution $\Upsilon(\chi,r,t)$ of higher-order modes is negligible in the limit of small self-propulsion forces, thus closing the hierarchy. 
Analogously to the previous approach, after integrating over the relative coordinate $r$, 
we obtain a continuity equation for the marginal density $\rho(\chi,t)$, i.e.,
\begin{equation}
    \partial_t \rho(\chi,t)= -\partial_\chi \left[ I(\chi,t) -\frac{D}{1+q}\partial_\chi \rho-v_w \rho \right],
    \label{density_eq_nosep}
\end{equation}
where
\begin{equation}
    \!\!I(\chi,t)=\int_{-\infty}^{\infty} \!\!\!dr \,\frac{ v_{\rm a}\left(\chi'\right) \sigma(\chi,r,t)}{(1+q)}
=\frac{\left<v_{\rm a}(\chi')\eta\,| \, \chi \right>}{1+q}\rho(\chi,t),
    \label{I_wave}
\end{equation}
and $\left<\cdot | \chi \right>$ denotes the conditional average at fixed $\chi$. We derive a close yet cumbersome analytical expression for $I(\chi,t)$ which  is  related to the local average swim speed  of the center of friction due to self-propulsion, see eq.~\eqref{I_wave} and sec.~4 of SM). Similarly, we also derive in the SM analytical expressions for  the stationary density and the flux in the comoving frame, which we use to analyze the directed transport in the regime of fast active traveling waves. In particular, for $D \tau_{\rm r} \ll \lambda^2$,
the average drift velocity $v_d$ reads
\begin{equation}
    \frac{v_d}{v_0}=\frac{l_p }{2\lambda(1+q)^2}\left[ \frac{\sin \psi_0}{|z_0|} + q \frac{\sin \psi_1}{|z_1|} \right],
    \label{vd_predicted}
\end{equation}
where we recall that $l_p = v_{0}\tau$ is the persistence length of the microswimmer, while $\psi_n$ and $|z_n|$ are the phase and the modulus, respectively, of the complex number 
\begin{equation}
    z_n=1 + \frac{\tau D}{\lambda^2(1+q)}+\frac{(1+q)\tau D}{q\ell^2} n +\mathrm{i}  \frac{\tau v_w}{\lambda},
\end{equation} 
where $\mathrm{i}$ is the imaginary unit. A general expression of the drift velocity for an arbitrary thermal diffusivity is given in sec.~4 of SM. 

Figure~\ref{drift_big} shows the behavior of the average drift $v_d$ as a function of the wave velocity $v_w$ in the regime $v_w>v_0$ of fast traveling waves. Unlike the case of $v_w<v_0$ (see fig.~\ref{drift_small}), the tactic behavior of the microswimmer does not exhibit a qualitative change as a function of the friction ratio $q$, with the drift occurring always along the direction of the active wave. 
However, as $q$ increases, this drift decreases because of the reduced mobility of the dimer. The drift velocity of the microswimmer attains its maximum value at a wave speed which scales as $v_w/v_0 \sim \lambda /l_p$. This can be qualitatively understood as following. Consider a single pulse of activity of spatial extent $\lambda$ travelling with a speed $v_w$. A microswimmer with its polarization against the direction of the travelling pulse will rapidly exit the pulse from the receding front. However, when the polarization is along the direction of the pulse, the microswimmer will be carried along with it until it switches its polarization which will cause it to exit the pulse. The optimum scenario corresponds to the condition $v_w \tau - v_0 \tau \sim \lambda$ in which the microswimmer effectively traverses the whole pulse before switching polarization. This results in a maximum of the drift speed at $v_w \sim \lambda /\tau$.

While the drift velocity of the dimer in fig.~\ref{drift_big} features a single peak, we find  both analytically and via numerical simulations that a second peak may appear at larger $v_w$, for large values of $q$ and persistence time $\tau$. The location of this additional peak depends on the spring relaxation time scale $\tau_{\rm r}$ but we defer a thorough investigation of its features and microscopic origin to future investigations. 

\section{Discussion}
\label{conclusions}

Our work shows that self-propelled cargo-carrying microswimmers interacting with a traveling wave of activity display a rich tactic behavior. Their response to such a wave is actually independent of the details of the activity, as evidenced by the equivalence of cargo carrying AOUPs and ABPs in terms of their coarse grained dynamics.
The tactic transition which emerges in the presence of slowly propagating waves relies on the possibility to control the preferential accumulation of the microswimmer in high/low activity regions, by tuning the friction of its cargo. In particular, we find a surfing effect when the directed migration along the activity wave is induced by an effective localization around the wave maxima. 
Considering, e.g., the experimental realization of Janus microswimmers as in ref.~\cite{lozano2019diffusing}, eq.~\eqref{q_threshold} implies $q_{\rm th} \simeq \kappa/(0.02 {\rm \,pN/\mu m})$ for $q_{\rm th}\gtrsim 1$. Accordingly, assuming for the cargo-carrier binding an elastic constant $\kappa \simeq 0.1 {\rm \,pN/\mu m}$, typical for soft matter, the tactic transition is predicted to occur at a cargo radius $\simeq 8\,{\rm \mu m}$, which is within experimental reach.
We speculate that a qualitatively similar tactic behavior may emerge spontaneously in a binary mixture of mutually attractive active and passive particles, upon formation of clusters of different sizes. 
It has  been recently shown that also molecules composed of two rigidly connected active particles~\cite{vuijk2022active} and dimers made of two active chiral particles~\cite{muzzeddu2022active} exhibit a transition in their effective localization in high/low activity regions. 
It will be interesting to study such active-matter systems subject to active traveling waves, and in the presence of external potentials~\cite{caprini2019entropy,garcia2021run}.

 We expect our predictions to have an impact on experimental studies on soft matter, biophysics, and nanotechnology.  Important  examples include cases in which synthetic Janus particles~\cite{baraban2012catalytic} and bacteria~\cite{akin2007bacteria} have been used  to efficiently transport and deliver microscopic objects  in specific target sites.    
 Moreover, our investigation could inspire future optimal design of existing {\em biohybrid} micromachines such as spermbots formed by assembling syntetic materials with sperm cells~\cite{medina2016cellular,magdanz2017spermatozoa}. The taxis transition unveiled by our minimal stochastic model may also have implications in biological processes at the microscale in which traveling waves play a key role, e.g., sound transduction in the cochlea~\cite{roberts1988hair,duke2003active} and signaling waves in cell development~\cite{di2022waves}.

\bibliographystyle{eplbib}
\bibliography{references}

\end{document}


\maketitle

\section{1. Deriving the mode equations}

In this section, we show how to derive the mode equations using the moment expansion technique~\cite{cates2013when, Solon2015comp,Adeleke2020}. We start the derivation from the Fokker-Planck equation (FPE) describing the evolution of $P(\bm{\chi},\bm{r},\bm{\eta},t)$ (eq.~(2) in the main text):
\begin{equation}
\begin{split}
    &\partial_t P(\bm{\chi},\bm{r},\bm{\eta},t)=-\nabla_{\bm{\chi}} \cdot \left[ -\bm{v}_w P +  \frac{1}{1+q}v_{\rm a}\left(\bm{\chi}'\right)\bm{\eta}P - \frac{D}{1+q}\nabla_{\bm{\chi}}P \right]+\\
    &\quad\quad\quad -\nabla_{\bm{r}}\cdot \left[ -\frac{1+q}{q\gamma} \nabla_{\bm{r}} UP+v_{\rm a}\left(\bm{\chi}'\right)\bm{\eta}P - \frac{1+q}{q}D \nabla_{\bm{r}} P \right]+\frac{1}{d\tau}\hat{\mathcal{L}}_{\bm{\eta}} P\,,
\end{split}
\end{equation}
where 
\begin{equation}
\bm{\chi}'=\bm{\chi}+q\bm{r}/(1+q), 
\label{Seq:chi-p}
\end{equation}
and the operator $\hat{\mathcal{L}}_{\bm{\eta}}$ is defined as 
\begin{equation}
    \hat{\mathcal{L}}_{\bm{\eta}}f(\bm{\eta})=\nabla^2_{\bm{\eta}}f(\bm{\eta})+d\nabla_{\bm{\eta}} \cdot \left[\bm{\eta} f(\bm{\eta}) \right]\,.
    \label{l_eta}
\end{equation}
We now expand the joint probability density as
\begin{equation}
    P(\bm{\chi},\bm{r},\bm{\eta},t)=\sum_{\bm{n}}\phi_{\bm{n}}(\bm{\chi},\bm{r},t)u_{\bm{n}}(\bm{\eta})\,,
    \label{decomposition}
\end{equation}
where $\bm{n} = \{n_1, n_2, \ldots, n_d \}$ is a set of non-negative integers, while $\left\{u_{\bm{n}}(\bm{\eta})\right\}$ is the corresponding set of eigenfunctions of the operator $\hat{\mathcal{L}}_{\bm{\eta}}$, given by
\begin{equation}
    u_{\bm{n}}(\bm{\eta})=\exp\left\{-\frac{d\bm{\eta}^2}{2} \right\}\prod_{i=1}^d H_{n_i}
    ( \sqrt{d}\,\eta_i);
    \label{un1}
\end{equation}
where $H_n(x)$ is the $n$-th Hermite polynomial in the probabilist convention~\cite{abramowitz1988handbook}. They satisfy the following eigenvalue equation 
\begin{equation}
    \hat{\mathcal{L}}_{\bm{\eta}}u_{\bm{n}}(\bm{\eta})=\lambda_{\bm{n}} u_{\bm{n}}(\bm{\eta}),
\end{equation}
where the eigenvalues $\lambda_{\bm{n}}$ are given by 
\begin{equation}
    \lambda_{\bm{n}}=-d\sum_{i=1}^d n_i.
\end{equation}
Moreover, it is convenient to introduce the family of functions $\left\{ \tilde{u}_{\bm{n}}(\bm{\eta})\right\}$ as
\begin{equation}
    \tilde{u}_{\bm{n}}(\bm{\eta})=(2 \pi)^{-d/2} \prod_{i=1}^d\frac{H_{n_i}
    ( \sqrt{d}\,\eta_i)}{n_i!},
\end{equation}
which are orthogonal to the eigenfunctions $\left\{u_{\bm{n}}(\bm{\eta})\right\}$, i.e.,
\begin{equation}
    \int d\bm{\eta}\, u_{\bm{n}}(\bm{\eta}) \tilde{u}_{\bm{m}}(\bm{\eta})=d^{-d/2} \delta_{\bm{n},\bm{m}},
\end{equation}
where $\delta_{\bm{n},\bm{m}} = \prod_{i=1}^d \delta_{n_i,m_i}$. Multiplying eq.~\eqref{decomposition} by $\tilde{u}_{\bm{0}}(\bm{\eta})$ and integrating over $\bm{\eta}$, we get
\begin{equation}
    \int d\bm{\eta} \, \tilde{u}_{\bm{0}}(\bm{\eta}) P(\bm{\chi},\bm{r},\bm{\eta},t) = \sum_{\bm{n}}\phi_{\bm{n}}(\bm{\chi},\bm{r},t) \int d\bm{\eta}\, \tilde{u}_{\bm{0}}(\bm{\eta}) u_{\bm{n}}(\bm{\eta})=d^{-d/2}\phi_{\bm{0}}(\bm{\chi},\bm{r},t)  \,,
    \label{phi_0}
\end{equation}
and after using the definition of $\tilde{u}_{\bm{0}}(\bm{\eta})$:
\begin{equation}
\begin{split}
    &\varphi(\bm{\chi},\bm{r},t)\equiv \int d\bm{\eta}\,P(\bm{\chi},\bm{r},\bm{\eta},t) =
        (2 \pi/d)^{d/2}
    \phi_{\bm{0}}(\bm{\chi},\bm{r},t) \,,
\end{split}
\label{rho_phi0}
\end{equation}
Accordingly, the first coefficient $\phi_{\bm{0}}(\bm{\chi},\bm{r},t)$ of the expansion in eq.~\eqref{decomposition} is related to the marginal density $\varphi(\bm{\chi},\bm{r},t)$. For later purposes, we recall that Hermite polynomials satisfy the recurrence relation \cite{abramowitz1988handbook}
\begin{equation}
    H_{n+1}(x)=xH_n(x)-H'_n(x),
    \label{sm:recH}
\end{equation}
and they form an Appell sequence, as they satisfy
\begin{equation}
    H'_n(x)=nH_{n-1}(x).
    \label{sm:appH}
\end{equation}
In order to lighten the notation, below we will denote by $\bm{n}_{\alpha \pm}$ the vector $(n_1,..,n_\alpha \pm1,...,n_d)$. Then, by using eqs.~\eqref{sm:recH} and \eqref{sm:appH} in eq.~\eqref{un1} one can write 
\begin{equation}
\begin{split}
    \eta_\alpha u_{\bm{n}}(\bm{\eta})
    &=\frac{1}{\sqrt{d}} \exp\left\{ -\frac{d\bm{\eta}^2}{2 } \right\}\sqrt{d}\eta_\alpha H_{n_\alpha}\left( \sqrt{d}\eta_\alpha\right) \prod_{\beta\neq \alpha} H_{n_\beta}\left( \sqrt{d}\eta_\beta \right)\\
    &=\frac{1}{\sqrt{d}}\exp\left\{ -\frac{d\bm{\eta}^2}{2} \right\}\left[H_{n_\alpha+1}\left( \sqrt{d}\eta_\alpha\right) + n_\alpha H_{n_\alpha-1}\left( \sqrt{d}\eta_\alpha\right) \right] \prod_{\beta\neq \alpha} H_{n_\beta}\left( \sqrt{d}\eta_\beta \right)\\
    &=\frac{1}{\sqrt{d}} u_{\bm{n}_{\alpha+}}(\bm{\eta}) +  \frac{n_\alpha}{\sqrt{d}} u_{\bm{n}_{\alpha-}}(\bm{\eta})\,.
\end{split}
\end{equation}
At this point we can project the FPE onto the $\left\{ \tilde{u}_{\bm{n}}(\bm{\eta})\right\}$ and obtain a set of equations for the coefficients $\left\{ \phi_{\bm{n}}(\bm{\chi},\bm{r},t)\right\}$. In the following, summation over repeated indices is implied.
For convenience, we will split the Fokker-Planck operator into the three contributions
\begin{equation}
    \partial_t P(\bm{\chi},\bm{r},\bm{\eta},t)=\left( \hat{\mathcal{L}}_{\bm{\chi}}+\hat{\mathcal{L}}_{\bm{r}}+\frac{1}{d\tau }\hat{\mathcal{L}}_{\bm{\eta}} \right)P,
    \label{S-eq:FP}
\end{equation}
where $\hat{\mathcal{L}}_{\bm{\eta}}$ is defined in eq.~\eqref{l_eta}, while
\begin{equation}
    \begin{split}
        &\hat{\mathcal{L}}_{\bm{\chi}}P=-\partial_\alpha \left[ \frac{1}{1+q}v_{\rm a}\left(\bm{\chi}+\frac{q}{1+q}\bm{r}\right)\eta_\alpha P -\frac{D}{1+q}\partial_\alpha P -v_w \delta_{\alpha,0} P\right],\\
        &\hat{\mathcal{L}}_{\bm{r}}P=-\partial'_\alpha \left[ -\frac{1+q}{q\gamma} \partial'_\alpha U(\bm{r})P+v_{\rm a}\left(\bm{\chi}+\frac{q}{1+q}\bm{r}\right)\eta_\alpha P - \frac{(1+q)}{q}D \partial'_\alpha P \right],
    \end{split}
    \label{S-eq:FP-proj}
\end{equation}
where we introduced the shorthand notation 
$\partial_\alpha \equiv \partial_{\chi_\alpha}$ and 
$\partial'_\alpha \equiv  \partial_{r_\alpha}$.
We separately project the various terms of the FPE onto $\tilde{u}_{\bm{m}}(\bm{\eta})$,  starting from its l.h.s.:
\begin{equation}
    \int d\bm{\eta} \,\tilde{u}_{\bm{m}}(\bm{\eta}) \partial_t P(\bm{\chi},\bm{r}, \bm{\eta},t)=\partial_t \phi_{\bm{n}}(\bm{\chi},\bm{r},t) \int d\bm{\eta} \, \tilde{u}_{\bm{m}}(\bm{\eta})u_{\bm{n}}(\bm{\eta})=d^{-d/2}\partial_t \phi_{\bm{m}}(\bm{\chi},\bm{r},t) \,.
\end{equation}
For the first term on the r.h.s., i.e., $\hat{\mathcal{L}}_{\bm{\chi}}P$, we have 
(for simplicity, we do not indicate below the dependence of $\phi_{\bm{n}}$ on $\bm{\chi}$ and $\bm{r}$):
\begin{equation}
    \begin{split}
        &\int d\bm{\eta} \,\tilde{u}_{\bm{m}}(\bm{\eta})\hat{\mathcal{L}}_{\bm{\chi}}P=\\
        &=-\partial_\alpha \left[ \frac{v_{\rm a}\left(\bm{\chi}'\right)\phi_{\bm{n}}}{1+q}\int d\bm{\eta} \,\tilde{u}_{\bm{m}}(\bm{\eta})\eta_\alpha u_{\bm{n}}(\bm{\eta}) -\left(\frac{D\partial_\alpha \phi_{\bm{n}}}{1+q} +v_w\delta_{\alpha,0} \phi_{\bm{n}} \right)\int d\bm{\eta} \, \tilde{u}_{\bm{m}}(\bm{\eta})u_{\bm{n}}(\bm{\eta})  \right]\\
        &=-\partial_\alpha \left\{ \frac{v_{\rm a}\left(\bm{\chi}'\right)\phi_{\bm{n}}}{\sqrt{d}(1+q)}\int d\bm{\eta} \,\tilde{u}_{\bm{m}}(\bm{\eta})\left[ u_{\bm{n}_{\alpha+}}(\bm{\eta}) + n_\alpha  u_{\bm{n}_{\alpha-}}(\bm{\eta}) \right] -\frac{Dd^{-d/2}}{1+q}\partial_\alpha \phi_{\bm{m}}-\frac{v_w\delta_{\alpha,0}}{d^{d/2}} \phi_{\bm{m}}\right\}\\
        &=-\partial_\alpha \left\{ \frac{d^{-(d+1)/2}}{1+q}v_{\rm a}\left(\bm{\chi}'\right)\left[\phi_{\bm{n}}\delta_{\bm{m},\bm{n}_{\alpha+}} + n_\alpha \phi_{\bm{n}}\delta_{\bm{m},\bm{n}_{\alpha-}}\right] -\frac{Dd^{-d/2}}{1+q}\partial_\alpha \phi_{\bm{m}} -v_w\delta_{\alpha,0}d^{-d/2} \phi_{\bm{m}} \right\}\\
        &=-\partial_\alpha \left\{ \frac{d^{-(d+1)/2}}{1+q}v_{\rm a}\left(\bm{\chi}'\right)\left[\phi_{\bm{m}_{\alpha-}} + (m_\alpha+1) \phi_{\bm{m}_{\alpha+}}\right] -\frac{Dd^{-d/2}}{1+q}\partial_\alpha \phi_{\bm{m}} -v_w\delta_{\alpha,0}d^{-d/2} \phi_{\bm{m}} \right\},
    \end{split}
    \label{Seq:FP-1}
\end{equation}
where we used $\delta_{\bm{m},\bm{n}_{\alpha-}}=\delta_{\bm{m}_{\alpha+},\bm{n}}$ and $\delta_{\bm{m},\bm{n}_{\alpha+}}=\delta_{\bm{m}_{\alpha-},\bm{n}}$. 
Similarly, the projection of the second term on the r.h.s.~of eq.~\eqref{S-eq:FP}, i.e., $\hat{\mathcal{L}}_{\bm{r}}P$, reads 
\begin{equation}
    \begin{split}
        &\int d\bm{\eta} \,\tilde{u}_{\bm{m}}(\bm{\eta})d^{d/2}\hat{\mathcal{L}}_{\bm{r}}P=\\
        &=-\partial'_\alpha \left\{ -\frac{(1+q)}{q\gamma} \partial'_\alpha U(\bm{r}) \phi_{\bm{m}} +d^{-1/2}v_{\rm a}\left(\bm{\chi}\right)\left[\phi_{\bm{m}_{\alpha-}} + (m_\alpha+1) \phi_{\bm{m}_{\alpha+}}\right] - \frac{(1+q)D}{q} \partial'_\alpha \phi_{\bm{m}} \right\}.
    \end{split}
    \label{Seq:FP-2}
\end{equation}
Finally, the last term in eq.~\eqref{S-eq:FP}, i.e., $\frac{1}{d\tau}\hat{\mathcal{L}}_{\bm{\eta}}P$, contributes as
\begin{equation}
    \begin{split}
        \frac{1}{d\tau}\int d\bm{\eta} \,\tilde{u}_{\bm{m}}(\bm{\eta})\hat{\mathcal{L}}_{\bm{\eta}}P=\frac{1}{d\tau}\phi_{\bm{n}}\int d\bm{\eta} \,\tilde{u}_{\bm{m}}(\bm{\eta})\hat{\mathcal{L}}_{\bm{\eta}} u_{\bm{n}}(\bm{\eta})=\frac{d^{-d/2-1}}{\tau}\lambda_{\bm{m}}\phi_{\bm{m}}.
    \end{split}
    \label{Seq:FP-3}
\end{equation}
Collecting the contributions in eqs.~\eqref{Seq:FP-1}, \eqref{Seq:FP-2}, and \eqref{Seq:FP-3}, the FPE projection onto $\tilde{u}_{\bm{m}}(\bm{\eta})$ yields the following set of coupled equations for the coefficients 
\begin{equation}
    \begin{split}
        &\partial_t \phi_{\bm{m}}(\bm{\chi},\bm{r},t)= \\
        &\quad -\partial_\alpha \left\{ \frac{1}{\sqrt{d}(1+q)}v_{\rm a}\left(\bm{\chi}'\right)\left[\phi_{\bm{m}_{\alpha-}} + (m_\alpha+1) \phi_{\bm{m}_{\alpha+}}\right] -\frac{D}{1+q}\partial_\alpha \phi_{\bm{m}} -v_w\delta_{\alpha,0} \phi_{\bm{m}} \right\} +\frac{\lambda_{\bm{m}}}{d\tau}\phi_{\bm{m}}\\
        &\quad -\partial'_\alpha \left\{ -\frac{(1+q)}{q\gamma} \partial'_\alpha U(\bm{r}) \phi_{\bm{m}} +\frac{v_{\rm a}\left(\bm{\chi}'\right)}{\sqrt{d}}\left[\phi_{\bm{m}_{\alpha-}} + (m_\alpha+1) \phi_{\bm{m}_{\alpha+}}\right] - \frac{(1+q)D}{q} \partial'_\alpha \phi_{\bm{m}} \right\}.
    \end{split}
    \label{set_eq_coeff}
\end{equation}
In particular, the dynamics of the first two modes $\varphi(\bm{\chi},\bm{r},t)$ and 
\begin{equation}
    \sigma_\alpha(\bm{\chi},\bm{r},t)\equiv \int d\bm{\eta}\,\eta_\alpha P(\bm{\chi},\bm{r},\bm{\eta},t)=\left( \frac{2\pi}{d}\right)^{d/2} \frac{\phi_{\bm{0}_{\alpha+}}(\bm{\chi},\bm{r},t)}{\sqrt{d}}\,,
\end{equation}
can be obtained by specialising eq.~\eqref{set_eq_coeff} to the cases $\bm{m}=\bm{0}$ and $\bm{m}=\bm{0}_{\alpha+}$, finding
\begin{equation}
\begin{split}
    &\partial_t \varphi(\bm{\chi},\bm{r},t)= -\partial_\alpha \left[-v_w \delta_{\alpha,0}\varphi + \frac{v_{\rm a}\left(\bm{\chi}'\right) \sigma_\alpha}{(1+q)} -\frac{D}{1+q}\partial_\alpha \varphi \right]\\
        &\quad\quad\quad-\partial'_\alpha \left[ -\frac{(1+q)}{q\gamma} \partial'_\alpha U \varphi +v_{\rm a}\left(\bm{\chi}'\right)\sigma_\alpha - \frac{(1+q)D}{q} \partial'_\alpha \varphi \right],
        \end{split}
        \label{density2variables}
\end{equation}
and
\begin{equation}
    \begin{split}
      &\partial_t \sigma_\alpha(\bm{\chi},\bm{r},t) = -\partial_\beta \left[ \frac{v_{\rm a}\left(\bm{\chi}'\right)
      \varphi \delta_{\alpha,\beta} 
      }{d(1+q)} -\frac{D}{1+q}\partial_\beta \sigma_\alpha -v_w \delta_{\beta,0}\sigma_\alpha \right]\\
        &\quad-\partial'_\beta \left[ -\frac{(1+q)}{q\gamma} \partial'_\beta U(\bm{r}) \sigma_\alpha +\frac{v_{\rm a}\left(\bm{\chi}'\right)
        \varphi \delta_{\alpha,\beta}
        }{d} - \frac{(1+q)D}{q} \partial'_\beta \sigma_\alpha \right]-\tau^{-1} \sigma_\alpha+\Upsilon(\bm{\chi},\bm{r},t),
    \end{split}
            \label{eq:sigma}
\end{equation}
with $\Upsilon(\bm{\chi},\bm{r},t)$ denoting the contributions due to higher-order modes. 
In order to simplify the notation, in the previous expression and in those which follow, the dependence on $(\bm{\chi},\bm{r},t)$ of $\varphi$ and $\sigma_\alpha$ is understood if not explicitly indicated.

In order to treat this hierarchy of equations, we adopt below two different approaches depending on the value of the phase velocity $v_w$ of the activity wave compared to the activity field $v_{\rm a}$ itself.


\section{2. Slow active traveling waves}

In the case of slowly propagating waves $v_w\ll v_0$, the hierarchy in eqs.~\eqref{density2variables} and \eqref{eq:sigma} can be closed by assuming that the activity field $v_{\rm a}$ varies on length scales much larger than the persistence length $l_p=\tau v_0$ (small gradients approximation), 
and considering quasi-stationary higher-order modes at time scales longer than $\tau$~\cite{cates2013when,Solon2015comp,Adeleke2020} 
Under these approximations, eq.~\eqref{eq:sigma} for the polarization field $\sigma_\alpha$ can be rewritten as
\begin{equation}
    \begin{split}
      \tau^{-1}\sigma_\alpha(\bm{\chi},\bm{r},t)= -\frac{\partial_\alpha \left[ v_{\rm a}\left(\bm{\chi}'\right)\varphi \right]}{(1+q)d}-\frac{\partial'_\alpha \left[v_{\rm a}\left(\bm{\chi}'\right)\varphi\right]}{d}+\frac{(1+q)}{q\gamma}\partial'_\beta \left[  \partial'_\beta U \sigma_\alpha \right] + \mathcal{O}(\partial^2)\,,
    \end{split}
    \label{qstat_pol}
\end{equation}
where $\mathcal{O}(\partial^2)$ denotes the contributions coming from higher-order powers of the gradient. 
This equation for $\sigma_\alpha(\bm{\chi},\bm{r},t)$ can be plugged into the continuity equation for the marginal density $\rho(\bm{\chi},t) = \int\! d\bm{r}\, \varphi(\bm{\chi},\bm{r},t)$, which can be obtained by integrating eq.~\eqref{density2variables} over the coordinate $\bm{r}$, finding
\begin{equation}
    \partial_t \rho(\bm{\chi},t)= -\partial_\alpha \left[\frac{1}{1+q}\int d\bm{r} \, v_{\rm a}\left(\bm{\chi}'\right) \sigma_\alpha(\bm{\chi},\bm{r},t) -v_w \delta_{\alpha,0}\rho(\bm{\chi},t) -\frac{D}{1+q}\partial_\alpha \rho(\bm{\chi},t) \right].
    \label{density1variable}
\end{equation}
On the r.h.s.~of this equation one recognizes the probability current $J_\alpha(\bm{\chi},t)$, corresponding to the expression in square brackets.
In addition to the diffusive term $\propto \bm{\nabla} \rho$ (with a renormalized diffusion coefficient $D/(1+q)$, as it refers to the diffusion of the center of friction) and to the current $\propto \bm{v}_w\rho$ due to the change of reference system, the additional contribution 
\begin{equation}
    I_\alpha(\bm{\chi},t) \equiv \frac{1}{1+q}\int d\bm{r} \, v_{\rm a}\left(\bm{\chi}'\right) \sigma_\alpha(\bm{\chi},\bm{r},t)
\label{Seq:addI}
\end{equation}
appears. By using eq.~\eqref{qstat_pol}, $I_\alpha(\bm{\chi},t)$ can be written as
\begin{equation}
    I_\alpha(\bm{\chi},t)=\frac{\tau}{1+q}\int d\bm{r} \, v_{\rm a}\left(\bm{\chi}'\right) \left\{ -\frac{\partial_\alpha \left[ v_{\rm a}\left(\bm{\chi}'\right)\varphi \right]}{d(1+q)}-\frac{\partial'_\alpha \left[v_{\rm a}\left(\bm{\chi}'\right)\varphi \right]}{d}-\frac{(1+q)}{q\gamma}\partial'_\beta \left[  F_\beta (\bm{r}) \sigma_\alpha \right] \right\},
\end{equation}
where $F_\beta(\bm{r})=-\partial_{\bm{r}_\beta} U(\bm{r})$. 
Integrating by part and using $\partial'_\alpha v_{\rm a}(\bm{\chi}')=\frac{q}{1+q}\partial_\alpha v_{\rm a}(\bm{\chi}')$ [which follows from eq.~\eqref{Seq:chi-p} and the definitions of $\partial'_\alpha$ and $\partial_\alpha$ given after eq.~\eqref{S-eq:FP-proj}], 
we can rewrite the previous expression as:
\begin{equation}
\begin{split}
    I_\alpha(\bm{\chi},t)&=\frac{\tau}{1+q}\int d\bm{r} \, v_{\rm a}\left(\bm{\chi}'\right) \left[ -\frac{\partial_\alpha \left[ v_{\rm a}\left(\bm{\chi}'\right)\varphi \right]}{d(1+q)}\right]\\
    &+\frac{\tau}{1+q}\int d\bm{r} \,
    \frac{q}{(1+q)d} \left[v_{\rm a}\left(\bm{\chi}'\right)\varphi \right]\partial_\alpha v_{\rm a}\left(\bm{\chi}'\right)
    \\
    & + \frac{\tau}{1+q} \frac{1}{\gamma} \int d\bm{r} \,
    F_\beta (\bm{r}) \sigma_\alpha 
    \partial_\beta v_{\rm a}\left(\bm{\chi}'\right). 
\end{split}
\label{flux1}
\end{equation}
We define now the quantity $\Sigma$ as
\begin{equation}
    \Sigma(\bm{\chi},t)\equiv \int d\bm{r} \, 
    F_\beta (\bm{r}) \sigma_\alpha (\bm{\chi},\bm{r},t)
    \partial_\beta v_{\rm a}\left(\bm{\chi}'\right) \,.
    \label{Seq:defSigma}
\end{equation}
By using eq.~\eqref{qstat_pol} into this expression for $\Sigma$ and by neglecting all terms 
$\mathcal{O}(\partial^2)$, we obtain:
\begin{equation}
\begin{split}
    \Sigma&=\tau \int d\bm{r} \,   F_\beta (\bm{r}) \partial_\beta v_{\rm a}\left(\bm{\chi}'\right) \left\{ -\frac{\partial'_\alpha \left[v_{\rm a}\left(\bm{\chi}'\right)\varphi\right]}{d}-\frac{(1+q)}{q\gamma}\partial'_\gamma \left[  F_{\gamma}(\bm{r}) \sigma_\alpha \right] \right\}\\
    &=\tau \int d\bm{r} \,    \left\{ \frac{v_{\rm a}\left(\bm{\chi}'\right)\varphi}{d}\partial'_\alpha \left[F_\beta (\bm{r}) \partial_\beta v_{\rm a}\left(\bm{\chi}'\right) \right]+\frac{(1+q)}{q\gamma}F_\gamma (\bm{r}) \sigma_\alpha \partial'_\gamma \left[ F_\beta (\bm{r}) \partial_\beta v_{\rm a}\left(\bm{\chi}'\right)   \right]  \right\}.
    \label{Sigma}
\end{split}
\end{equation}
The last line can be further simplified  by considering separately 
\begin{equation}
\begin{split}
    \partial'_\alpha \left[F_\beta (\bm{r}) \partial_\beta v_{\rm a}\left(\bm{\chi}'\right) \right]&=\partial'_\alpha F_\beta (\bm{r}) \partial_\beta v_{\rm a}\left(\bm{\chi}'\right)+F_\beta (\bm{r}) \partial'_\alpha \partial_\beta v_{\rm a}\left(\bm{\chi}'\right)\\
    &=\partial'_\alpha F_\beta (\bm{r}) \partial_\beta v_{\rm a}\left(\bm{\chi}'\right)+\frac{q}{1+q}F_\beta (\bm{r}) \partial_\alpha \partial_\beta v_{\rm a}\left(\bm{\chi}'\right)\\
    &=\partial'_\alpha F_\beta (\bm{r}) \partial_\beta v_{\rm a}\left(\bm{\chi}'\right)+\mathcal{O}(\partial^2)\,.
    \end{split}
    \label{Seq:simp-1}
\end{equation}
Moreover, since the interaction potential is modeled by a spring with stiffness $\kappa$ and zero rest length, we have that
\begin{equation}
    \partial'_\alpha F_\beta (\bm{r})=-\kappa \delta_{\alpha,\beta}=\partial'_\beta F_\alpha (\bm{r}).
\end{equation}
Accordingly, eq.~\eqref{Seq:simp-1} can be written as
\begin{equation}
    \partial'_\alpha \left[F_\beta (\bm{r}) \partial_\beta v_{\rm a}\left(\bm{\chi}'\right) \right] \simeq  -\kappa \delta_{\alpha,\beta} \partial_\beta v_{\rm a}\left(\bm{\chi}'\right)=-\kappa \partial_\alpha v_{\rm a}\left(\bm{\chi}'\right),
\end{equation}
and thus eq.~\eqref{Sigma} becomes
%
%
%
\begin{equation}
    \begin{split}
        \Sigma&=-\kappa \tau \int d\bm{r} \,    \left[\frac{v_{\rm a}\left(\bm{\chi}'\right)\varphi}{d}\partial_\alpha v_{\rm a} \left( \bm{\chi}'\right) +\frac{(1+q)}{q\gamma}F_\gamma (\bm{r}) \sigma_\alpha \partial_\gamma v_{\rm a} \left( \bm{\chi}'\right)\right]=\\
        &=-\frac{\kappa \tau}{2d} \int d\bm{r} \, \varphi \partial_\alpha v^2_{\rm a}\left(\bm{\chi}'\right) -\kappa \tau \frac{(1+q)}{q\gamma}\Sigma,
  \end{split}
  \end{equation}      
where, in the last line, we used the  
definition of $\Sigma$, see eq.~\eqref{Seq:defSigma}. Accordingly, the previous equation can be solved and yields 
\begin{equation}
        \Sigma(\bm{\chi},t)  =-\frac{1}{2d}\frac{\gamma \tau/\tau_{\rm r} }{ 1 + \frac{1+q}{q}\frac{\tau}{\tau_{\rm r}}} \int d\bm{r} \, \varphi(\bm{\chi},\bm{r},t) \partial_\alpha v^2_{\rm a}\left(\bm{\chi}'\right),
    \label{I_fin}
\end{equation}
where, as in the main text, we introduced $\tau_{\rm r} = \gamma/\kappa$.
%
%
%

This expression of $\Sigma(\bm{\chi},t)$ can be used in eq.~\eqref{flux1}, finding
\begin{equation}
\begin{split}
    I_\alpha(\bm{\chi},t)&=-\frac{\tau}{d(1+q)^2}\int d\bm{r} \,  v^2_{\rm a}\left(\bm{\chi}'\right)\partial_\alpha \varphi  
    -\frac{1}{2}\frac{\tau}{d(1+q)^2}
    \epsilon
    \int d\bm{r} \,\varphi \partial_\alpha v^2_{\rm a}\left(\bm{\chi}'\right),
\end{split}
\end{equation}
where we introduced the tactic coupling
\begin{equation}
    \epsilon=1-\frac{q}{1+\frac{1+q}{q}\frac{\tau}{\tau_{\rm r}}},
    \label{tactic_coupling}
\end{equation}
reported in eq.~(8) 
of the main text.

Moreover, if the typical distance between the active carrier and the cargo is small compared to the persistence length, we can approximate $\varphi = \varphi(\bm{\chi},\bm{r},t)$ in the integrands above as:
\begin{equation}
    \varphi(\bm{\chi},\bm{r},t) \approx \rho(\bm{\chi},t)\delta(\bm{r})\,.
\end{equation}
Within this approximation, the total current $J_{\alpha}(\bm{\chi},t)$ introduced after 
eq.~\eqref{density1variable} can be written as
\begin{equation}
    J_{\alpha}(\bm{\chi},t)=V_{\rm eff,\alpha}(\bm{\chi})\rho(\bm{\chi},t)-\partial_{\alpha}\left[D_{\rm eff}\rho(\bm{\chi},t)\right]\,,
\end{equation}
where the effective drift and diffusivity are, respectively, given by
\begin{equation}
   V_{\rm eff,\alpha}(\bm{\chi})=  (1-\epsilon/2)\partial_\alpha D_{\rm eff}(\bm{\chi}) -v_w\delta_{\alpha,0} \quad\mbox{and}\quad 
   D_{\rm eff}(\bm{\chi})=\frac{D}{1+q}+\frac{\tau v^2_{\rm a}(\bm{\chi})}{d(1+q)^2},
\label{Seq:drift-diff}
\end{equation}
which are reported in eqs.~(6) and (7) of the main text. 
%
%
%
%
The stationary solution of the effective Fokker-Planck equation
\begin{equation}
    \partial_t \rho(\bm{\chi},t)=-\nabla_{\bm{\chi}}\cdot \left[\bm{V}_{\rm eff}(\bm{\chi})\rho(\bm{\chi},t) - \nabla_{\bm{\chi}}(D_{\rm eff}(\bm{\chi})\rho(\bm{\chi},t)) \right]\,
\end{equation}
can be easily proved to be~\cite{hanggi1990reaction, goel2016stochastic,merlitz2018linear}:
\begin{equation}
    \frac{\rho(\bm{\chi})}{\rho_b}=\frac{L\, D_{\rm eff}^{-1}(\chi_0) \int_0^L dx\exp \left\{-\int_{\chi_0}^{\chi_0+x}dy\frac{V_{\rm eff,0}(y)}{D_{\rm eff}(y)}  \right\}}{\int_0^L\,du\int_0^Ldx\, D_{\rm eff}^{-1}(u)\exp \left\{-\int_{u}^{u+x}dy\frac{V_{\rm eff,0}(y)}{D_{\rm eff}(y)}  \right\}},
    \label{steady_state_density_supp}
\end{equation}
in the case of periodic boundary conditions, as reported in the main text. Moreover, the system can sustain a finite stationary flux in the comoving frame
\begin{equation}
    J_0=\frac{ \rho_b L \left[ 1-\exp \left\{-\int_{0}^{L}dy\frac{V_{\rm eff,0}(y)}{D_{\rm eff}(y)}  \right\}\right]}{\int_0^L\,du\int_0^Ldx\, D_{\rm eff}^{-1}(u)\exp \left\{-\int_{u}^{u+x}dy\frac{V_{\rm eff,0}(y)}{D_{\rm eff}(y)}  \right\}}
    \label{stat_flux_comoving}
\end{equation}
in the direction $\bm{e}_0$, which can be used in order to compute the average drift velocity $v_d=J_0/\rho_b+v_w$ in the lab frame, which is reported in eq.~(13)  of the main text.

\section{3. Drift velocity at $q_{\rm th}$}
In this section we show that the drift velocity $v_d$ [see eq.~(13) 
in the main text], derived in the limit of slow propagating activity fields $v_w\ll v_0$, vanishes if $q$ takes the threshold value $q_{\rm th}$ reported in eq.~(10) 
of the main text.
In particular, when $q=q_{\rm th}$, the tactic coupling $\epsilon$ in eq.~\eqref{tactic_coupling} [alternatively, see eq.~(8) of the main text] 
vanishes and the effective drift in eq.~\eqref{Seq:drift-diff} becomes
\begin{equation}
    V_{\rm eff,\alpha}(\bm{\chi})=\partial_\alpha D_{\rm eff}(\bm{\chi}) -v_w\delta_{\alpha,0}\,.
\end{equation}
This expression can be used in eq.~\eqref{stat_flux_comoving} in order to calculate the stationary current in the comoving frame. In particular, the denominator of that expression reads:
\begin{equation}
    \begin{split}
    &\int_0^L\!du\int_0^L\!dx\, D_{\rm eff}^{-1}(u)\exp \left\{-\int_{u}^{u+x}dy \, \left[\partial_\alpha \ln D_{\rm eff}(y) -\frac{v_w}{D_{\rm eff}(y)}  \right]\right\}\\
    &\quad\quad= \int_0^L\!du\int_0^L\!dx\, D_{\rm eff}^{-1}(u+x)\exp \left\{\int_{u}^{u+x}\!dy \, \frac{v_w}{D_{\rm eff}(y)}  \right\}\\
    &\quad\quad= \frac{1}{v_w}\int_0^L\!du \left[\exp \left\{\int_{u}^{u+L}dy \, \frac{v_w}{D_{\rm eff}(y)}\right\} -1 \right] \\
    &\quad\quad= \frac{L}{v_w}\left[\exp \left\{\int_{0}^{L}\!dy \, \frac{v_w}{D_{\rm eff}(y)}\right\} -1 \right],
    \end{split}
    \label{Seq:num}
\end{equation}
where in the last equality we used that the effective drift $D_{\rm eff}(y)$ is a periodic function with period $L$. Analogously the numerator of eq. \eqref{stat_flux_comoving} is given by:
\begin{equation}
    \begin{split}
        \rho_b L \left[ 1-\exp \left\{-\int_{0}^{L}dy\frac{\partial_\alpha D_{\rm eff}(y) -v_w}{D_{\rm eff}(y)}  \right\}\right]=\rho_b L \left[ 1-\exp \left\{\int_{0}^{L}dy\frac{v_w}{D_{\rm eff}(y)}  \right\}\right].
    \end{split}
    \label{Seq:den}
\end{equation}
Combining eqs.~\eqref{Seq:num} and \eqref{Seq:den}, the average drift velocity $v_d$, given after eq.~\eqref{stat_flux_comoving}, reads:
\begin{equation}
    v_d=\frac{J_0}{\rho_b}+v_w = v_w \frac{\rho_b L \left[ 1-\exp \left\{\int_{0}^{L}dy\frac{v_w}{D_{\rm eff}(y)}  \right\}\right]}{\rho_b L\left[\exp \left\{\int_{0}^{L}dy \, \frac{v_w}{D_{\rm eff}(y)}\right\} -1 \right]}+v_w=0
\end{equation}


\section{4. Fast active traveling waves}
In this section we derive analytical expressions for the stationary density, stationary current and average drift velocity in the regime of fast active traveling waves, i.e., for $v_w\gg v_0$. To this aim, we adopt a different strategy to close the hierarchy of equations governing the dynamics of the modes given by Eqs \eqref{density2variables} and \eqref{eq:sigma}, which hinges on assuming a small activity $v_{0}$ compared to the wave velocity $v_w$. 
For simplicity, we present the derivation for the one-dimensional case $d=1$ with 
the sinusoidal activity field reported in eq.~(11) of the main text.
%
%
%
The extension to the case with $d\neq 1$ is straightforward.
To implement the new closure scheme, we start from the dynamics of the polarisation field given by eq.~\eqref{eq:sigma}, which can be conveniently rewritten as:
\begin{equation}
    \hat{\mathcal{L}}_{\sigma}\sigma(\chi,r,t)=-\frac{\partial_\chi \left[ v_{\rm a}(\chi')\varphi \right]}{(1+q)}-\partial_r \left[ v_{\rm a}(\chi')\varphi \right]+\Upsilon(\chi,r,t)\,,
    \label{eq:sigma2}
\end{equation}
where $\varphi = \varphi(\chi,r,t)$, the position $\chi'$ of the active carrier in the comoving frame is defined as in eq.~\eqref{Seq:chi-p}, specialized to $d=1$, and 
$\Upsilon(\chi,r,t)$ includes the contributions of 
higher-order modes. In the previous equation, the operator $\hat{\mathcal{L}}_{\sigma}$ is defined as
\begin{equation}
\hat{\mathcal{L}}_{\sigma}=\partial_t + \frac{1}{\tau} -v_w\partial_\chi -\frac{D}{1+q} \partial^2_\chi - \frac{(1+q)D}{q}\left[\partial^2_r + \frac{1}{\ell^2}\partial_r r\right],
\label{op_sigma}
\end{equation}
with the characteristic length 
$\ell=\sqrt{D \tau_{\rm r}}$ and $\tau_{\rm r}=\gamma/k$. We first determine the Green function $G(\chi,r,t;\chi_0,r_0,t_0)$ of the operator $\hat{\mathcal{L}}_{\sigma}$, defined as:
\begin{equation}
    \hat{\mathcal{L}}_{\sigma} G(\chi,r,t;\chi_0,r_0,t_0)=\delta(\chi-\chi_0)\delta(r-r_0)\delta(t-t_0)\,.
    \label{green_dimer}
\end{equation}
Note that, due to the translational invariance of the operator $\hat{\mathcal{L}}_{\sigma}$ in the variables $\chi$ and $t$, one expects $G(\chi,r,t;\chi_0,r_0,t_0)$ to be a function of $\chi-\chi_0$ and $t-t_0$. The presence  of the interparticle potential, instead, breaks the transaltional invariance of $\hat{\mathcal{L}}_{\sigma}$ with respect to $r$ and therefore $G(\chi,r,t;\chi_0,r_0,t_0)$ depends separately on $r$ and $r_0$. Accordingly, we can write $G(\chi,r,t;\chi_0,r_0,t_0) = G(\chi-\chi_0,r,t-t_0;0,r_0,0) \equiv G(\chi-\chi_0,r,t-t_0;r_0)$ where in the last equality we introduce a convenient shorthand notation. The function $G(\chi,r,t;r_0)$
can be conveniently determined
by expanding it in the Fourier-Hermite basis
\begin{equation}
    G(\chi,r,t;r_0)=\frac{1}{\ell}\sum_{n=0}^{\infty} \int \frac{d \omega}{2 \pi} \int \frac{d \tilde{q}}{2 \pi} \tilde{G}_n(\tilde{q},\omega;r_0) e^{i\tilde{q}\chi+i\omega t} u_n(r),
    \label{Seq:Gf-exp}
\end{equation}
where $u_n(r)$ is given by
\begin{equation}
    u_{n}(r)= 
    e^{-r^2/(2\ell^2)}
     H_{n}(r/\ell)\,,
     \label{eq:u}
\end{equation}
and $H_n(x)$ is the $n$-th probabilist's Hermite polynomial.
With this expansion, the l.h.s.~of eq.~\eqref{green_dimer} becomes
\begin{equation}
    \frac{1}{\ell}\sum_{n=0}^{\infty} \int \frac{d \omega}{2\pi}\frac{d\tilde{q}}{2 \pi}  \left[i\omega + \tau^{-1} +\frac{D}{1+q} \tilde{q}^2 +iv_w \tilde{q}  + \frac{(1+q)D}{q}\frac{n}{\ell^2} \right] \tilde{G}_n(\tilde{q},\omega;r_0) e^{i\tilde{q}\chi+i\omega t} u_n(r),
\label{Seq:Glhs}
\end{equation}
while its r.h.s.~is
\begin{equation}
    \frac{1}{\ell}\sum_{n=0}^{\infty} \int \frac{d \omega}{2 \pi} \int \frac{d \tilde{q}}{2 \pi} \tilde{u}_n(r_0) e^{i\tilde{q}\chi+i\omega t} u_n(r),
    \label{Seq:Grhs}
\end{equation}
where we used the fact that $\delta(r-r_0)$ in eq.~\eqref{green_dimer} can be written as
\begin{equation}
    \frac{1}{\ell}\sum_{n=0}^\infty \tilde{u}_n(r_0) u_n(r)=\delta(r-r_0)\,,
\end{equation}
and the functions $\tilde{u}_n(r)$ are defined as:
\begin{equation}
    \tilde{u}_n(r)=\frac{1}{\sqrt{2\pi}n!}H_n(r/\ell).
    \label{eq:utilde}
\end{equation}
Accordingly, by comparing eq.~\eqref{Seq:Glhs} with eq.~\eqref{Seq:Grhs}, the Green function in reciprocal space turns out to be given by
\begin{equation}
    \tilde{G}_n(\tilde{q},\omega;r_0)=\frac{\tilde{u}_n(r_0)}{i\omega + \tau^{-1} +\frac{D}{1+q} \tilde{q}^2 +iv_w \tilde{q}  + \frac{(1+q)D}{q}\frac{n}{\ell^2}}.
\end{equation}
After inserting this expression of $\tilde{G}_n(\tilde{q},\omega;r_0)$ into eq.~\eqref{Seq:Gf-exp}, one can readily calculate the integral in $\omega$ via the residue theorem. The corresponding residue is a Gaussian function of $\tilde q$ and thus the corresponding integral is also straightforward, with the final result
%
%
%
%
 \begin{equation}
     G(\chi,r,t;r_0)
    =\Theta(t) \frac{\exp\left\{- \frac{t}{\tau}-\frac{\left(\chi +v_w t \right)^2}{4  \frac{D}{1+q} t} \right\}}{\sqrt{4 \pi \frac{D}{1+q} t}}\frac{1}{\ell}\sum_{n=0}^{\infty} \exp\left\{-\frac{(1+q)Dn}{q\ell^2}t \right\} \tilde{u}_n(r_0)u_n(r),
    \label{Seq:G-exp-sum}
\end{equation}
where the Heaviside function $\Theta$ is defined such that $\Theta(t>0) =1$ and $\Theta(t\le 0) =0$.
Before considering the last summation, we introduce the quantity 
\begin{equation}
    s=\exp\left\{-\frac{(1+q)}{q}\frac{Dt}{\ell^2} \right\}<1.
    \label{Seq-def-s}
\end{equation}
In terms of $s$ the remaining sum in eq.~\eqref{Seq:G-exp-sum} can be written as
\begin{equation}
\begin{split}
    \frac{1}{\ell}\sum_{n=0}^{\infty} s^n \tilde{u}_n(r_0)u_n(r)&=\frac{1}{\ell \sqrt{2 \pi}} \exp\left\{-\frac{r^2}{2\ell^2} \right\}\sum_{n=0}^{\infty} \frac{s^n}{ n!} H_n\left(\frac{r_0}{\ell} \right)H_n\left(\frac{r}{\ell} \right)\\
    &=\frac{1}{\sqrt{2 \pi \ell^2 (1-s^2)}} \exp\left\{-\frac{(r-sr_0)^2}{2(1-s^2)\ell^2} \right\}
    \end{split}
    \label{Seq-sum-M}
\end{equation}
where we used the expression of $u_n$ and $\tilde u_n$ given in eqs.~\eqref{eq:u} and \eqref{eq:utilde}, respectively,
and, in the second equality, we used Mehler's formula~\cite{abramowitz1988handbook} for 
probabilist's 
Hermite polynomials, i.e.,
\begin{equation}
    \sum_{n=0}^{\infty} \frac{s^n}{ n!} H_n(x)H_n(y)=\frac{1}{\sqrt{1-s^2}} \exp \left\{-\frac{s^2(x^2+y^2)-2sxy}{2(1-s^2)} \right\} \quad\mbox{for}\quad -1<s<1.
\end{equation}
Accordingly, by using eqs.~\eqref{Seq:G-exp-sum}, \eqref{Seq-def-s}, and \eqref{Seq-sum-M} the 
the Green function in eq.~\eqref{green_dimer} reads:
\begin{equation}
    G(\chi,r,t;r_0)=\Theta(t)\exp\left\{-t/\tau\right\}  \frac{\exp\left\{-\frac{\left(\chi +v_w t\right)^2}{4Dt/(1+q)} \right\}}{\sqrt{4 \pi Dt/(1+q)}} \frac{\exp\left\{-\frac{(r-sr_0)^2}{2(1-s^2)\ell^2} \right\}}{\sqrt{2 \pi \ell^2 (1-s^2)}}.
\end{equation}
Once this Green function is known, one can determine $\sigma(\chi,r,t)$ by computing the convolution integral over $\chi_0$, $t_0$, and $r_0$ of the product between $G(\chi-\chi_0,r,t-t_0;r_0)$ and the r.h.s.~of eq.~\eqref{eq:sigma2} evaluated for $\chi=\chi_0$, $r=r_0$, and $t=t_0$.
Once $\sigma(\chi,r,t)$ is known, we can calculate the current contribution to eq.~\eqref{density1variable} given by eq.~\eqref{Seq:addI}, specialized to the case $d=1$. 
In particular, one has
\begin{equation}
    \begin{split}
        &I(\chi,t)=\\
        &-\int dr \, v_{\rm a}\left(\chi'\right) \int_{-\infty}^{\infty} \!\!d\chi_0 dr_0 dt_0\, G(\chi-\chi_0,r,t-t_0;r_0)\frac{\partial_{\chi_0} \left[ v_{\rm a}(\chi_0+\frac{qr_0}{1+q})\varphi(\chi_0,r_0,t_0) \right]}{(1+q)^2}\\
        &-\int dr \, v_{\rm a}\left(\chi'\right) \int_{-\infty}^{\infty} \!\!d\chi_0 dr_0 dt_0\, G(\chi-\chi_0,r,t-t_0;r_0) \frac{\partial_{r_0} \left[ v_{\rm a}(\chi_0+\frac{qr_0}{1+q})\varphi(\chi_0,r_0,t_0) \right]}{(1+q)}\\
        &+\int dr \, v_{\rm a}\left(\chi'\right) \int_{-\infty}^{\infty} \!\!d\chi_0 dr_0 dt_0\, G(\chi-\chi_0,r,t-t_0;r_0)\frac{\Upsilon(\chi_0,r_0,t_0)}{(1+q)}.
    \end{split}
\end{equation}
The latter integral can be computed under the approximation of small activity field compared to $v_w$, 
by keeping only terms of the lowest order in $v_0$. 
For this reason, we neglect the contribution coming from higher-order modes $\Upsilon(\chi,r,t)$, 
thus closing the hierarchy, and we evaluate the first two integrals by assuming that the density field
\begin{equation}
    \varphi(\chi_0,r_0,t_0) = \rho_b \frac{e^{-r_0^2/(2\ell^2)}}{\sqrt{2 \pi \ell^2}} + \mathcal{O}(v_0/v_w)
    \label{Seq:phi-app}
\end{equation}
is approximately equal to the one in equilibrium, i.e., for $v_{\rm a}\propto v_0 =0$, and $\rho_b$ is the bulk density.
In this way, all integrals appearing in the first two lines are standard Gaussian integrals, and can be easily calculated. As a result, $I(\chi,t)$ is actually independent of time (as $\varphi$ in eq.~\eqref{Seq:phi-app}) and is given by
\begin{equation}
    \begin{split}
        I(\chi)&=-\frac{\rho_b \tau v_0^2 e^{-\frac{q^2\ell^2}{2\lambda^2(1+q)^2}}}{\lambda (1+q)^2} \Bigg\{\frac{\cos (\chi/\lambda + \psi_0)}{|z_0|}-q \frac{\cos \left(\chi/\lambda \right)}{\left(1+\frac{(1+q)\tau D}{q\ell^2}\right)}\\
        &\quad\quad\quad+e^{-\frac{q^2\ell^2}{2\lambda^2(1+q)^2}}\sum_{n=0}^\infty \frac{\left[\frac{ q^2 \ell^2}{\lambda^2(1+q)^2} \right]^n}{n!}\left[ f_n(\chi)+qf_{n+1}(\chi)\right]\Bigg\},
    \end{split}
    \label{Seq:Ichifast}
\end{equation}
with
\begin{equation}
    f_n(\chi)=\frac{(-1)^n \sin(2 \chi/\lambda+\psi_n)-\sin \psi_n}{2|z_n|}\,,
\end{equation}
and where $\psi_n$ and $|z_n|$ are the phase and the modulus, respectively, of the complex number $z_n$ defined in eq.~(19) of the main text. 
%
%
%
%
In order to compute the marginal probability density $\rho(\chi)$ in the steady state, we impose that the probability current in eq.~\eqref{density1variable} equals the constant $J$. 
Accordingly, one has to solve the following differential equation, 
\begin{equation}
     \frac{D}{1+q}\partial_\chi \rho(\chi)+v_w \rho(\chi)=I(\chi)-J,
     \label{diff_eq_rho}
\end{equation}
with $I(\chi)$ given in eq.~\eqref{Seq:Ichifast}. 
This can be done by first computing the Green function $G_1$, defined by
\begin{equation}
     \left(\frac{D}{1+q}\partial_\chi +v_w\right) G_1(\chi-\chi_0)=\delta(\chi-\chi_0),
\end{equation}
which reads (in the case of $v_w>0$)
\begin{equation}
    G_1(\chi-\chi_0)=\frac{(1+q)}{D} \Theta(\chi-\chi_0)\exp\left\{-\frac{(1+q)v_w}{D}(\chi-\chi_0) \right\},
\end{equation}
and then the following convolution:
\begin{equation}
    \rho(\chi)=\frac{(1+q)}{D} \int_{-\infty}^{\chi}d\chi'\,\exp\left\{-\frac{(1+q)v_w}{D}(\chi-\chi') \right\} I(\chi')-\frac{J}{v_w} .
\end{equation}
The contribution coming from the homogeneous solution of eq. \eqref{diff_eq_rho} vanishes under periodic boundary conditions.
Also in this case, the convolution involves Gaussian integrals, the standard calculation of which is not reported here for the sake of space. As a result, the stationary density $\rho(\chi)$ can be expressed as
\begin{equation}
    \begin{split}
            \rho(\chi)=-\frac{\rho_b \tau  v_0^2}{D\lambda(1+q)}e^{-\frac{q^2\ell^2}{2\lambda^2(1+q)^2}} \Bigg[&\frac{\cos (\chi/\lambda + \psi_0+\varphi(\lambda))}{|\zeta(\lambda)||z_0|}-q \frac{\cos \left(\chi/\lambda + \varphi(\lambda) \right)}{|\zeta(\lambda)|\left(1+\frac{(1+q)\tau D}{q\ell^2}\right)}+ \\&+e^{-\frac{q^2\ell^2}{2\lambda^2(1+q)^2}}\sum_{n=0}^\infty \frac{\left(\frac{ q^2 \ell^2}{\lambda^2(1+q)^2} \right)^n}{n!}\left[ g_n(\chi)+qg_{n+1}(\chi)\right]\Bigg] -\frac{J}{v_w},
    \end{split}
\end{equation}
where the functions $g_n(\chi)$ are defined as
\begin{equation}
    g_n(\chi)=\frac{(-1)^n \sin(2\chi/\lambda+\psi_n+\varphi(\lambda/2))}{2|\zeta(\lambda/2)||z_n|}-\frac{\sin(\psi_n)}{2\frac{(1+q)v_w}{D}|z_n|},
\end{equation}
and where $\varphi(\lambda)$ and $|\zeta(\lambda)|$ are the phase and the modulus, respectively, of the $\lambda$-dependent complex number
\begin{equation}
\begin{split}
    \zeta(\lambda)=\frac{(1+q) v_w}{D}-\mathrm{i}\lambda^{-1}.
\end{split}
\end{equation}
Moreover, by imposing the normalization of the marginal density $\rho(\chi)$, we find the expression of the stationary current in the comoving frame $J$:
\begin{equation}
    J=-\frac{v_w}{L} \left[1- \frac{\tau v_0^2}{2v_w\lambda(1+q)^2}e^{-\frac{q^2\ell^2}{\lambda^2(1+q)^2}}\sum_{n=0}^\infty \frac{\left(\frac{ q^2 \ell^2}{\lambda^2(1+q)^2} \right)^n}{n!} \left(\frac{\sin(\psi_n)}{|z_n|} + \frac{q\sin(\psi_{n+1})}{|z_{n+1}|}\right) \right]\,,
\end{equation}
and, as a consequence, the average drift velocity:
\begin{equation}
    \frac{v_d}{v_0}=  \frac{l_p}{2\lambda(1+q)^2}e^{-\frac{q^2\ell^2}{\lambda^2(1+q)^2}}\sum_{n=0}^\infty \frac{\left(\frac{ q^2 \ell^2}{\lambda^2(1+q)^2} \right)^n}{n!} \left(\frac{\sin(\psi_n)}{|z_n|} + \frac{q\sin(\psi_{n+1})}{|z_{n+1}|}\right).
\end{equation}
The previous equation is reported in the main text (see eq. (18)) in the limit of small thermal diffusivity $D\tau_{\rm r}\ll\lambda^2$.

\bibliographystyle{eplbib}
\bibliography{references}